\documentclass[conference]{IEEEtran}
\usepackage[pdftex]{graphicx}

\usepackage[caption=false,font=footnotesize]{subfig}

\usepackage{amsmath}

\usepackage{cite}

\usepackage{url}

\usepackage{siunitx}

\usepackage{lipsum}

\usepackage[mathscr]{euscript}

\usepackage{xcolor,colortbl}
\usepackage{color,soul}

\definecolor{morange}{rgb}{0.8,0.2,0}
\definecolor{mblue}{rgb}{0,0.3,1.0}
\definecolor{mpink}{rgb}{1.0,0.6,0.6}
\definecolor{mgreen}{rgb}{0.1,0.6,0.2}
\definecolor{mgoodgreen}{rgb}{0.9,1.0,0.7}

\definecolor{Gray}{gray}{0.85}

\usepackage{multirow}
\usepackage{array}
\newcolumntype{L}[1]{>{\raggedright\let\newline\\\arraybackslash\hspace{0pt}}m{#1}}
\newcolumntype{C}[1]{>{\centering\let\newline\\\arraybackslash\hspace{0pt}}m{#1}}
\newcolumntype{R}[1]{>{\raggedleft\let\newline\\\arraybackslash\hspace{0pt}}m{#1}}

\newcolumntype{G}{>{\columncolor{mgoodgreen}}c}

\newcommand{\xth}[1]{ {#1}\textrm{th} }

\newcommand{\tend}{t_{\text{end}}}

\newcommand{\txn}[1]{\text{Tx}_{#1}}
\newcommand{\rxn}[1]{\text{Rx}_{#1}}

\newcommand{\nrx}[1]{N^{\rxn{#1}}}
\newcommand{\nrxk}[2]{N^{\rxn{#1}}[#2]}
\newcommand{\bitTxik}[2]{s^{\txn{#1}}_{#2}}
\newcommand{\bitRxik}[2]{s^{\rxn{#1}}_{#2}}

\newcommand{\Fij}[1]{F_\text{#1}}
\newcommand{\Fijk}[2]{F_\text{#1}[#2]}

\newcommand{\probb}{\mathbf{P}}
\newcommand{\probbeGiven}[1]{\probb_{e|#1}}
\newcommand{\probbe}{\probb_{e}}

\newcommand{\isiw}{\eta}

\DeclareMathOperator*{\argminn}{\arg\min}

\begin{document}

 \title{Machine Learning based Channel Modeling for Molecular MIMO Communications}


\author{
\IEEEauthorblockN{Changmin Lee, H. Birkan Yilmaz, and Chan-Byoung Chae}
\IEEEauthorblockA{School of Integrated Technology\\ 
Yonsei University, Korea\\ 
Email:\{cm.lee, birkan.yilmaz, cbchae\}@yonsei.ac.kr}
\and
\IEEEauthorblockN{Nariman Farsad and Andrea Goldsmith}
\IEEEauthorblockA{Department of Electrical Engineering \\
Stanford University, USA\\
Email: \{nfarsad,andrea\}@stanford.edu }
}


\maketitle

\begin{abstract}
In diffusion-based molecular communication, information particles locomote via a diffusion process, characterized by random movement and heavy tail distribution for the random arrival time. As a result, the molecular communication shows lower transmission rates. To compensate for such low rates, researchers have recently proposed the molecular multiple-input multiple-output (MIMO) technique. Although channel models exist for single-input single-output (SISO) systems for some simple environments, extending the results to multiple molecular emitters complicates the modeling process. In this paper, we introduce a technique for modeling the molecular MIMO channel and confirm the effectiveness via numerical studies.\\

\end{abstract}


\IEEEpeerreviewmaketitle

\section{Introduction}
Molecular communication (MC) is quite different from traditional electromagnetic wave communication in that MC conveys information by utilizing molecules. Communication on such a small scale is a challenging task~\cite{farsad2016comprehensiveSO}. Recently, researchers have been focusing on MC due to its potential for enabling complex applications of nano-technology that require collaboration of very small entities. For example, MC can be utilized in nano-scale communications for nano-robots' communication and for health-monitoring issues~\cite{farsad2016comprehensiveSO,nakano2013molecularC_BOOK,akyildiz2011nanonetworksAN}. 

Compared to traditional electromagnetic wave-based communication, MC in nano-scale communication has three advantages. First, low energy consumption is one of the important needs of nano-scale communication and MC shows better performance (in terms of energy consumption) than do traditional methods. Second, MC is able to decrease the size of the antenna-- validated by many examples in nature~\cite{farsad2016comprehensiveSO,nakano2013molecularC_BOOK,akyildiz2011nanonetworksAN,guo2015molecularVE}. 
Finally, the biocompatible characteristics for \textit{in vivo} systems demonstrated by MC are greater than those of traditional communication~\cite{farsad2016comprehensiveSO,nakano2013molecularC_BOOK,akyildiz2011nanonetworksAN}, as long as harmless molecules are used.

In diffusion-based MC, the received signal is determined by the received molecules and the information particles move randomly via a diffusion process,  which shows random movement. 
One of the main challenges in MC is to develop valid models for representing the received signal in different environments. Studies on MC, such as modulation, detection, and receiver design utilize the channel models that depend on the environment conditions~\cite{arjmandi2013diffusionBN,noe2014optimalRD,hsieh2013asynchronousIE,kilinc2013receiverDF_JSAC,azadi2016novelEM}. Previous studies have mostly focused on the molecular single-input single-output (SISO) communication system. Indeed, few studies have focused on the molecular multiple-input multiple-output (MIMO) communication system. In~\cite{srinivas2012molecularCI_inverseG}, the authors presented an analytical channel model for an absorbing receiver in a ${\mbox{1-dimensional}}$ (1D) environment. In~\cite{kadlor2012molecularCU_drift_TNBS}, the authors enhanced the channel model by considering the flow effect. In~\cite{yilmaz2014threeDC}, the authors presented a theoretical molecular  channel model for a 3D environment. The model is based on a point transmitter and an absorbing spherical receiver, where the expected cumulative number of received molecules is formulated with respect to time. In~\cite{akkaya2015effectOR_receptor_COML}, the channel model from~\cite{yilmaz2014threeDC} has been enhanced by adding the receptor effect. When we consider a molecular MIMO channel with two absorbing antennas, analytical derivation becomes intractable. Therefore, semi analytical methods are utilized in which the channel taps are acquired from simulations and plugged into analytical derivations. In~\cite{koo2016molecularMIMO_JSAC}, the authors presented a molecular MIMO channel model using simulation data. What is known to be a hard problem is calculating an exact channel model for the molecular MIMO. Therefore, we need a more practical method for acquiring the channel taps without simulation. 

Modeling the molecular MIMO channel with absorbing receiver antennas is still an open issue. In this paper, we propose a machine learning based approach to model a $2\times2$ molecular MIMO channel with point transmitter antennas and absorbing receiver antennas. We utilize an artificial neural network (ANN) based channel modeling technique. We develop two different techniques that utilize the ANN method-- \emph{one-machine} and \emph{two-machines} techniques. The differences between the two is one thing investigated here. Then, we utilize the model of the proposed technique for bit error rate (BER) calculations. 

\section{System Model}
\begin{figure}[!t]
	\begin{center}
		\includegraphics[width=0.9\columnwidth,keepaspectratio]%
		{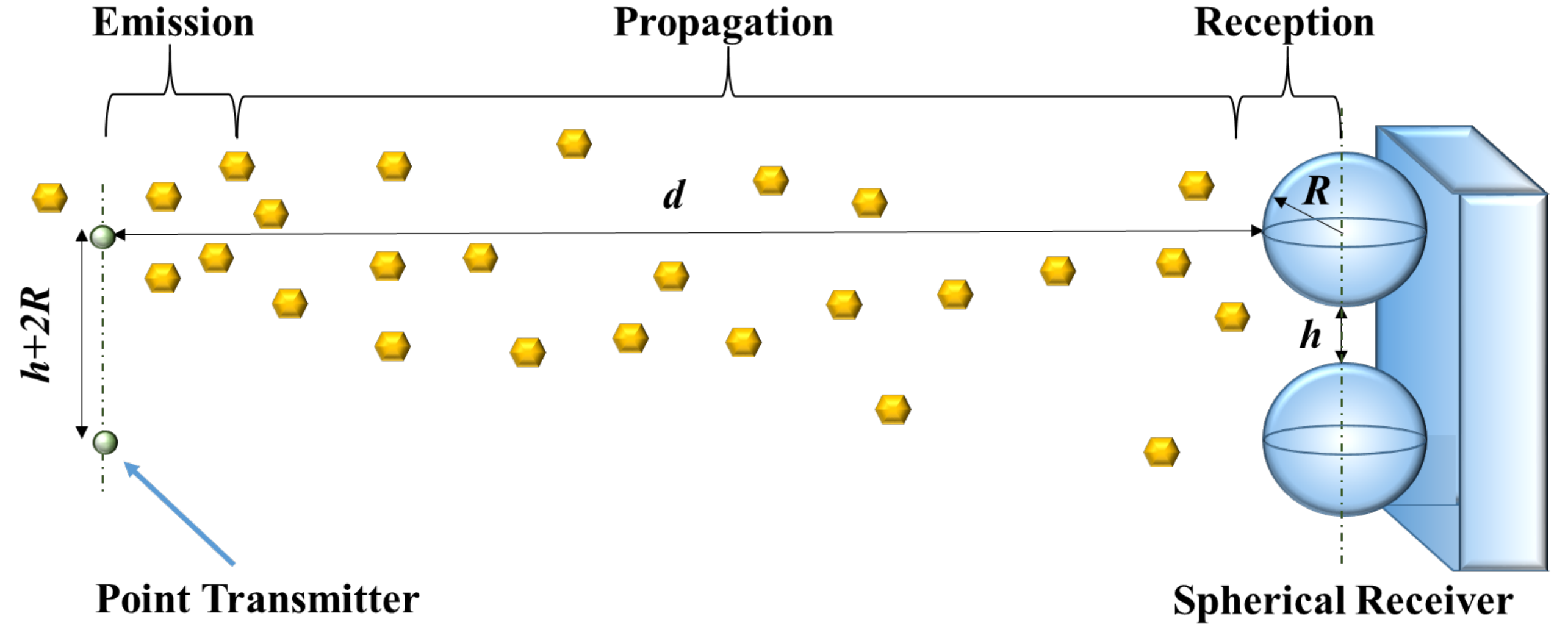}
		\caption{The system model of the $2\times 2$ molecular MIMO communication with two point transmitters and two spherical receivers.}
		\label{fig_system_modell}
	\end{center}
	\vspace {-20pt}
\end{figure}

Consider a $2\times 2$ molecular MIMO system, as depicted in Fig.~\ref{fig_system_modell}. It shows two point transmitters and two spherical receivers attached to a cuboid body. In this model, we assume perfect alignment of antennas. Transmitter antennas are separated by distance~$h$, which also holds for the receiver antennas. The radius of a spherical receiver antenna is denoted by~$R$. The distance between corresponding transmitter and receiver antennas is denoted by~$d$. 

When molecules are emitted from the point transmitter antennas, they propagate in the 3D environment by the diffusion process. The diiffusion process is characterized by the diffusion coefficient~$D$. Receiver antennas absorb the molecules that hit the receiver surface and each receiver antenna counts the number of received molecules. Absorbed molecules are counted only once (i.e., molecules contribute to the molecular signal only once). We assume that if a molecule hits the cuboid body, then the molecule is reflected to the medium.

In this paper, we utilize the SISO channel model to develop a molecular MIMO channel model. In~\cite{yilmaz2014threeDC}, the authors presented and analyzed expectations of SISO channel response is presented and analyzed from the perspective of channel characteristics. The authors presented the formula for the fraction of molecules that hit the receiver until time $t$, as follows:
\begin{align}
\begin{split}
F_\text{hit}^{3\text{D}} (t)=  \frac{R}{d\!+\!R} \,\text{erfc} \left( \frac{d}{\sqrt{4Dt\,}}\right) 
\end{split}
\label{eqn_3d_frac_received_point_src}
\end{align}
where $d$ and $\text{erfc}(.)$ represent the distance and the complementary error function. When there is only one spherical receiver, there is a circular symmetry, all the points at the same radius are equivalent and the solution for the system of differential equations is enabled. However, this model cannot define the molecular MIMO channel  exactly. Hence we use the above SISO channel model with additional adjustment parameters for each receiver antenna.

\begin{figure*}[!t]
	\begin{center}
		\includegraphics[width=1.99\columnwidth,keepaspectratio]%
		{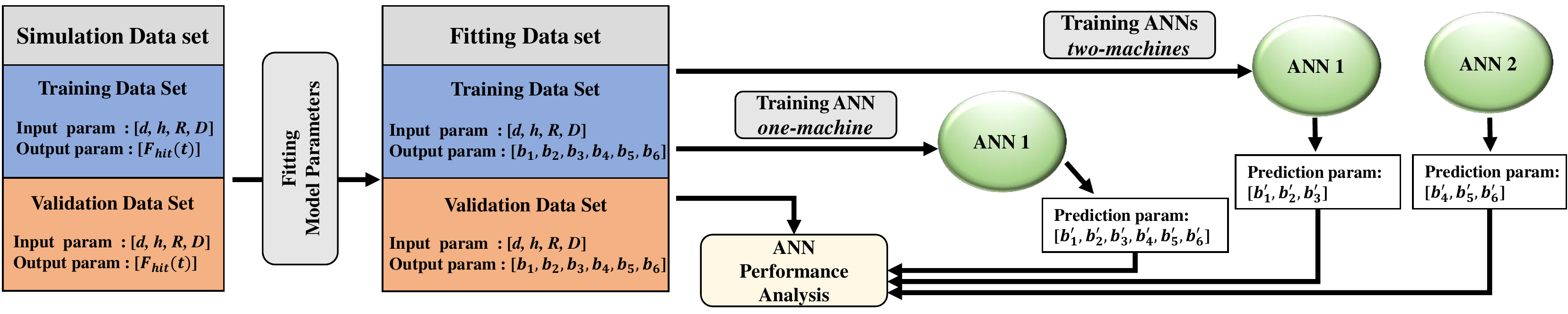}
		\caption{The flowchart of the proposed techniques: \emph{one-machine}  and \emph{two-machines}. From left to right, the process 1 deals with fitting model parameters by utilizing the dataset from the simulator; the process 2 deals with training ANNs on the training dataset that is obtained from the process 1. The output from process 1 consists of input-output pairs where the input is ${(d,\,h,\,R,\,D)}$ and the output is the model parameters (i.e., $b_i$'s).}
		\label{fig_proposed_technq_flowchart}
	\end{center}
	\vspace{-15pt}
\end{figure*}

\section{Proposed Technique for Channel Modeling}
Finding model parameters for the ${2\times 2}$ molecular MIMO channel defines all the required data for evaluating the channel impulse response. In this communication system, there are four emission cases: only transmitter 1 ($\txn{1}$) or 2 ($\txn{2}$) emits, two transmitters emit simultaneously, and none of them emit molecules. Due to the rectangular symmetry of the topology, we can model the molecular MIMO channel by considering the emission of a single transmitter. In this case, we can formulate all four cases by just using the channel response of the single emitter case. Therefore, by using the molecular MIMO communication simulator, we collected data for the channel response of receiver 1 ($\rxn{1}$) and 2 ($\rxn{2}$)\ when a single antenna emitted molecules.

The proposed technique consists of two processes: fitting the channel model parameters and learning the patterns in the input-output dataset by ANNs. After the training phase, the trained ANN is able to predict the channel model parameters effectively without the need of simulations. A representative scheme of the proposed technique is depicted in  Fig.~\ref{fig_proposed_technq_flowchart}.

\subsection{Channel Model and Fitting}
For fitting the simulation data, we use two different model functions. Because $\rxn{1}$'s and $\rxn{2}$'s responses exhibit different schemes due to the distance difference, we use the corresponding distances in model functions. We consider the case that only the $\txn{1}$ emits molecules for analyzing the cumulative channel functions at the $\rxn{1}$ (i.e., $\Fij{11}(\cdot)$) and $\rxn{2}$ (i.e., $\Fij{21}(\cdot)$). We use three scaling factors for the model function, which is given as follows:
\begin{align}
\begin{split}
\Fij{11} (t, b_1, b_2, b_3)=  b_1 \, \frac{R}{d\!+\!R} \,\text{erfc} \left( \frac{d}{(4D)^{b_2} \, t^{b_3}}\right)
\end{split}
\label{eqn_model_rx1}
\end{align}
where $b_1$, $b_2$, and $b_3$ represent the model fitting parameters. These model-fitting parameters are introduced so as to compensate for the discrepancy between the SISO and MIMO models. Similarly we define the response at the $\rxn{2}$ (due to the cross link interference) as follows:
\begin{align}
\begin{split}
\Fij{21} (t, b_4, b_5, b_6)=  b_4 \, \frac{R}{\sqrt{d^2\!+\!h^2}\!+\!R} \,\text{erfc} \left( \frac{\sqrt{d^2\!+\!h^2}}{(4D)^{b_5} \, t^{b_6}}\right) 
\end{split}
\label{eqn_model_rx2}
\end{align}
where $b_4$, $b_5$, and $b_6$ are model fitting parameters. 

To find the model parameters, we use the nonlinear least squares curve-fitting technique. Assuming that we have $N$ observations during the simulation, we formulate the parameter estimation problem with $m$ parameters as follows:
\begin{align}
\begin{split}
\argminn\limits_{b_1,...,b_m} \sum\limits_{k=1}^{N} \left(\Fij{ij} (t_k, b_1,...,b_m)- S_{ij}^{3\text{D}} (t_k) \right)^2
\end{split}
\label{eqn_fitting_lsq}
\end{align}
where $S_{ij}^{3\text{D}} (t)$ corresponds to the mean simulation data that is representing the ratio of hitting molecules until time $t$. Please note that for each receiver we use the corresponding model function from \eqref{eqn_model_rx1} or \eqref{eqn_model_rx2}.

The output of the curve-fitting process consists of the model parameters. Hence, we obtain model parameters for each simulation case, which forms the dataset of the next process. This dataset structure is depicted in  Fig.~\ref{fig_proposed_technq_flowchart} (i.e., the model parameters in the fitting data set blocks).

\subsection{Training Artificial Neural Network}
We propose to model the channel response in the diffusion-based $2\times 2$ molecular MIMO communication system with two point transmitters and two spherical receivers by using a machine-learning technique. We used one of the popular machine-learning techniques-- ANN. ANN has simple and neuron-like nodes with thresholds and the connections with weights. Basically, the thresholds and the weights are adjusted by back-propagation (which evaluates the gradient and takes a step accordingly to minimize the loss function) for learning the input-output pattern. 

The dataset is divided into two disjointed subsets as training (TDS) and validation (VDS) datasets. TDS is utilized for training the ANN to get the desired output for given inputs. For the ANN training, we used back-propagation with Bayesian regularization, which updates the weights and bias values according to Levenberg-Marquardt optimization. Bayesian regularization helps to minimize a combination of squared errors and weights together. Hence, it helps to determine the ANN parameters that generalize the pattern in the input-output pairs without overlearning. After the learning phase, we utilize the trained ANN to estimate the channel parameters for different cases. Note that a trained ANN does not require any simulation data, i.e., the required inputs are the system parameters such as $d$, $h$, $R$, and $D$.

Furthermore, we considered the effect of the number of ANNs for the modeling purpose. Hence, to realize a machine learning technique, we considered two different methods-- \emph{one-machine} and \emph{two-machines} (Fig.~\ref{fig_proposed_technq_flowchart}). First, to  train a single ANN, we used the input parameters of $d$, $h$, $R$, $D$ and considered all the output parameters as model parameters (i.e., $b_1$, $b_2$, $b_3$, $b_4$, $b_5$, $b_6$). Second, for two ANNs, we used the input parameters of $d$, $h$, $R$, $D$ in a similar way but we considered the output parameters separately. The first ANN's output parameters were $b_1$, $b_2$, $b_3$ and the second ANN's were $b_4$, $b_5$, $b_6$. The first ANN predicts the channel model parameters of $\rxn{1}$, which is aligned with the emitting transmitter and the second ANN predicts the channel model parameters of $\rxn{2}$. For both techniques, we used the same number of hidden layer nodes, i.e., \emph{one-machine} technique's ANN consists of 30 hidden layer nodes, so each ANN of the \emph{two-machines} technique uses 15 hidden layer nodes.

\subsection{Using ANN Output for Theoretical BER Evaluation}
For a case study, we utilized the ANN results to evaluate the BER of a selected case. In this molecular MIMO communication system, information is sent using a sequence of symbols that are spread over sequential time slots ($t_s$). We considered binary concentration shift keying (BCSK) modulation at each antennas, i.e., $N$ molecules are emitted for a bit-1 and no emission is done for a bit-0~\cite{kuran2011modulationTF_ICC,nr13}. Demodulation takes place at the end of each symbol slot and BCSK symbols are demodulated by thresholding the number of received molecules at $\rxn{1}$ ($\nrx{1}$) and $\rxn{2}$ ($\nrx{2}$) in a given symbol slot. The demodulated symbol at the $\xth{m}$ symbol slot of $\rxn{i}$ is denoted by $\bitRxik{i}{m}$. In MC, the number of received molecules is highly affected by previous emissions, which causes inter-symbol-interference (ISI). For tractability, we consider a finite number of symbol slots for ISI and we neglect the insignificant part of the ISI. With a given ISI window ($\isiw$), the number of received molecules at the $\xth{m}$ symbol slot is the sum of Binomial distributed random variables where each summand represents the number of hitting molecules due to a previous emission. For tractability, if we use Gaussian approximation for Binomial random variables and if we consider an ISI window of $\isiw$, then the general case is formulated as follows:
\begin{align}
\begin{split}
\nrxk{i}{m} &\sim \mathscr{N}(\mu,\, \sigma^2 )\\
\mu &= \sum\limits_{k=0}^{\isiw} N \bitTxik{i}{m\!-\!k}\,\, \Fijk{ii}{k} +\sum\limits_{k=0}^{\isiw} N \bitTxik{j}{m\!-\!k}\,\, \Fijk{ij}{k}\\
\sigma^2 &= \sum\limits_{k=0}^{\isiw} N \bitTxik{i}{m\!-\!k}\,\, \Fijk{ii}{k}\,(1\!-\!\Fijk{ii}{k})\\
		&+ \sum\limits_{k=0}^{\isiw} N \bitTxik{j}{m\!-\!k}\,\, \Fijk{ij}{k}\,(1-\Fijk{ij}{k})
\end{split}
\label{eqn_nrx}
\end{align}
where $\mathscr{N}(\mu,\, \sigma^2 )$, $\nrxk{i}{m}$, $\bitTxik{i}{k}$, and $\Fijk{ij}{k}$ denote a Gaussian random variable with mean $\mu$ and variance $\sigma^2$, the number of received molecules for $\rxn{i}$ at the $\xth{m}$ symbol slot, the bit value for $\txn{i}$ to send at the $\xth{k}$ symbol slot, and probability of hitting to $\rxn{i}$ for the molecules emitted from $\txn{j}$ at the $\xth{k}$ following symbol slot. Please note that $\Fijk{ij}{0}$ corresponds to the hitting probability at the current symbol slot. By using the output of the ANN technique, we are able to model $\nrxk{i}{m}$, which enables the evaluation of BER analytically as follows:
\begin{align}
\probbe = \sum\limits_{\bitTxik{1}{m\!-\!\eta : m},\, \bitTxik{2}{m\!-\!\eta : m}} \hspace{-0.5cm} \probbeGiven{\bitTxik{1}{m\!-\!\eta : m},\, \bitTxik{2}{m\!-\!\eta : m}} \probb(\bitTxik{1}{m\!-\!\eta : m}) \probb(\bitTxik{2}{m\!-\!\eta : m}) 
\label{eq_ber}
\end{align}
where
\begin{align}
\begin{split}
\probbeGiven{\bitTxik{1}{m\!-\!\eta : m},\, \bitTxik{2}{m\!-\!\eta : m}} &=\frac{1}{2} \probb(\bitRxik{1}{m} \neq \bitTxik{1}{m}|\bitTxik{1}{m\!-\!\eta : m},\, \bitTxik{2}{m\!-\!\eta : m}) \\
 	&+\frac{1}{2} \probb(\bitRxik{2}{m} \neq \bitTxik{2}{m}|\bitTxik{1}{m\!-\!\eta : m},\, \bitTxik{2}{m\!-\!\eta : m}).
\end{split}
\end{align}
These probabilities are evaluated by the tail probabilities of the Gaussian random variable that is defined in \eqref{eqn_nrx}.


\section{Results and Analysis}

Common system parameters for simulations, TDS, and VDS are presented in Table~\ref{tbl_system_parameters}. From the given datasets, each of the VDS and TDS have ${5\times 3\times 2\times 3\!=\!90}$ different cases, making a total of $180$ cases. Each simulation case is replicated $500$ times to estimate the mean channel model.

\subsection{Received Signal Analysis}
\begin{figure}[!t]
\begin{center}
	\includegraphics[width=0.9\columnwidth,keepaspectratio]%
	{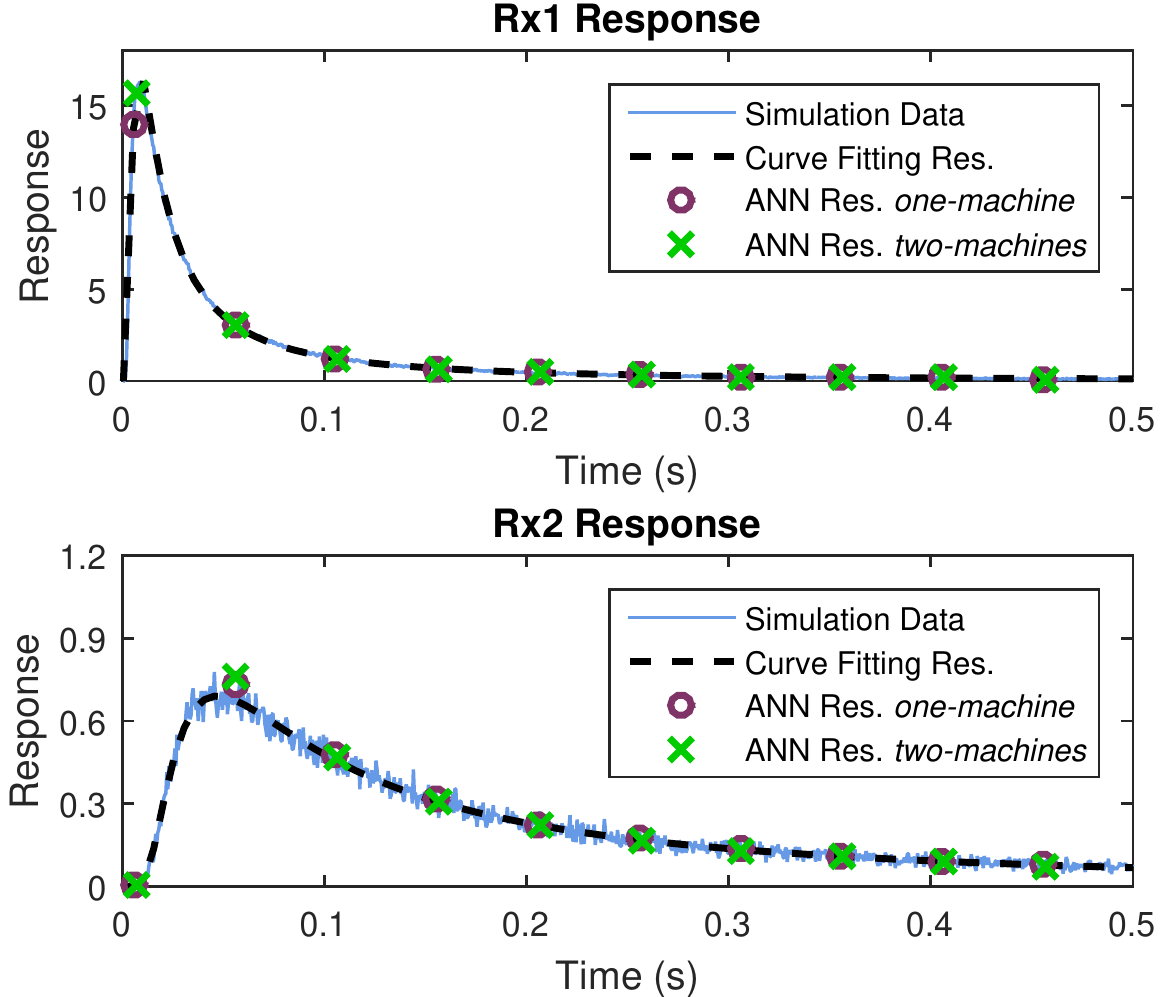}
	\caption{The received signal plots for $d\!=\!\SI{2}{\micro\meter}$, $h\!=\!\SI{1}{\micro\meter}$, $R\!=\!\SI{3}{\micro\meter}$, $D\!=\!\SI{100}{\micro\meter^2/\second}$ with the time resolution of $\SI{1}{\milli\second}$. $F_\text{11}$ is shown on the top response plot and $F_\text{21}$ is shown on the bottom plot.}
	\label{fig____received_signal1}
\end{center}
\end{figure}

\begin{figure}[!t]
\begin{center}
	\includegraphics[width=0.9\columnwidth,keepaspectratio]%
	{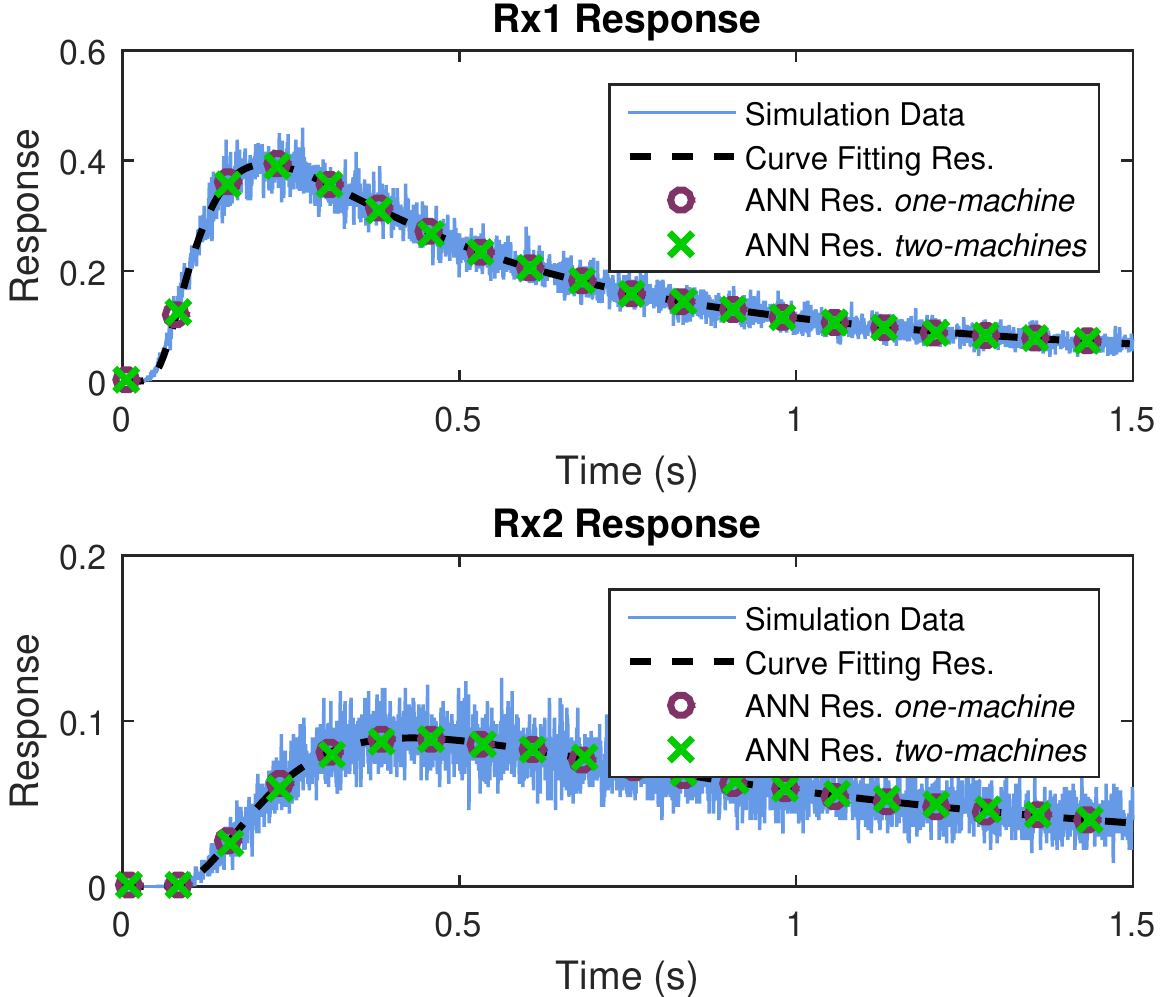}
	\caption{The received signal plots for $d\!=\!\SI{8}{\micro\meter}$, $h\!=\!\SI{1}{\micro\meter}$, $R\!=\!\SI{5}{\micro\meter}$, $D\!=\!\SI{50}{\micro\meter^2/\second}$ with the time resolution of $\SI{1}{\milli\second}$. $F_\text{11}$ is shown on the top response plot and $F_\text{21}$ is shown on the bottom plot.}
	\label{fig____received_signal2}
\end{center}
\end{figure}

In Figs.~\ref{fig____received_signal1} and~\ref{fig____received_signal2}, the received signals are plotted for simulation data, curve-fitting, and ANN (\emph{one-machine} and \emph{two-machines}) techniques. Note that the ANN technique requires no simulation data, while the curve-fitting method does. After training an ANN, we estimated channel model parameters for VDS by giving only the system parameters as input.

It can be seen from Fig.~\ref{fig____received_signal1} and~\ref{fig____received_signal2} that the trained ANN estimates the channel model parameters effectively for ${d=\SI{2}{\micro\meter}}$ and ${d=\SI{8}{\micro\meter}}$ cases. We observe that the trained ANNs perform better at longer distances, a finding that will be detailed and quantified in the following subsection. 

\begin{table}[t]
\begin{center}
\caption{Range of parameters used in the experiments}
\renewcommand{\arraystretch}{1.14}
\label{tbl_system_parameters}
\begin{tabular}{p{5.2cm} l}
\hline
\bfseries{Parameter} 							& \bfseries{Value} \\ 
\hline 
Number of emitted molecules			& $3\,000$\\
Duration of channel simulations  ($\tend$) 		& $\SI{1.5}{\second}$\\ 
Replication 						& $500$\\
TDS Distances ($d$) 				& $\{3,\, 5,\, 7,\, 9,\, 11\}\,\, \si{\micro\metre} $\\
VDS Distances ($d$) 				& $\{2,\, 4,\, 6,\, 8,\, 10\}\,\, \si{\micro\metre} $\\
Distance between antennas ($h$)		& $\{0,\, 1,\, 2\}\,\,\si{\micro\metre}$\\
Diffusion coefficients ($D$) 	& $\{50,\, 100\}\,\,\si{\micro\metre^2/\second}$\\
Receiver radius ($R$)			& $\{3,\, 5,\, 7\}\,\,\si{\micro\metre}$\\
\hline
\end{tabular} 
\end{center}
\renewcommand{\arraystretch}{1}
\vspace{-15pt}
\end{table}

\subsection{RMSE Analysis}
To analyze the performance of \emph{one-machine} and \emph{two-machines}, we evaluated the root mean square error (RMSE) of cases with respect to simulation data in terms of the channel response. In Tables~\ref{table_RMSE} and~\ref{table_RMSE_half data}, the mean RMSEs of the channel responses for the given methods are presented for different TDS sizes. In the tables, the cases are grouped by the number of ANNs, $d$ (distance between transmitter and receiver), and $D$ (diffusion coefficient). 

In Table~\ref{table_RMSE}, mean RMSE values are given for different groups of input parameters. The main takeaway of these results is that the \emph{one-machine} method learns the channel model parameters slightly better than the \emph{two-machines} method. A second takeaway is that the proposed techniques perform better when we increase the distance. For the short distances, the compensation parameters cannot work effectively enough to model the molecular MIMO channel parameters. The gain of the proposed technique is higher for the longer distances both in terms of RMSE (i.e., the modeling performance) and the run-time (simulations for the longer distances  require much more run-time).
\renewcommand*{\arraystretch}{1.17}
\setlength\tabcolsep{2.5pt} 
\begin{table}[t]
\caption{Mean RMSE values for ANN techniques with full TDS}
\begin{center}
\begin{tabular}{| C{1.35cm} | C{0.8cm} | c | c || c | c |}
\cline{1-6}
\multicolumn{2}{| c |}{} & \multicolumn{2}{ c ||}{ $D=\SI{50}{\micro\meter^2/\second}$} & \multicolumn{2}{ c| }{$D=\SI{100}{\micro\meter^2/\second}$} \\ \cline{3-6}
\multicolumn{2}{| c |}{} & One ANN  &  Two ANN 	&  One ANN  &  Two ANN	\\ \hline
\vspace{-0.4cm}$d=\SI{2}{\micro\meter}$  & \vspace{-0.32cm} $F_{11}(t)$	&  0.1561 &  0.1596 &  0.3043 &    0.2827 	\\ \cline{3-6}
                                         & \vspace{-0.32cm} $F_{21}(t)$ &  0.0437 &  0.0458 &  0.0537 &    0.0540 	\\ \cline{1-6}
\vspace{-0.4cm}$d=\SI{4}{\micro\meter}$  & \vspace{-0.32cm} $F_{11}(t)$	&  0.0574 &  0.0575 &  0.0922 &    0.0924 	\\ \cline{3-6}
                                         & \vspace{-0.32cm} $F_{21}(t)$	&  0.0195 &  0.0201 &  0.0229 &    0.0234 	\\ \cline{1-6}
\vspace{-0.4cm}$d=\SI{6}{\micro\meter}$  & \vspace{-0.32cm} $F_{11}(t)$ & 0.0395 &  0.0395 &  0.0538 &    0.0537 	\\ \cline{3-6}
                                         & \vspace{-0.32cm} $F_{21}(t)$	&  0.0176 &  0.0177 &  0.0217 &   0.0218 	\\ \cline{1-6}
\vspace{-0.4cm}$d=\SI{8}{\micro\meter}$  & \vspace{-0.32cm} $F_{11}(t)$	&  0.0305 &  0.0305 &  0.0383 &    0.0383 	\\ \cline{3-6}
                                         & \vspace{-0.32cm} $F_{21}(t)$	&  0.0152 &  0.0153 &  0.0198 &    0.0199 	\\ \cline{1-6}
\vspace{-0.4cm}$d=\SI{10}{\micro\meter}$ & \vspace{-0.32cm} $F_{11}(t)$	&  0.0239 &  0.0239 &  0.0303 &   0.0304 	\\ \cline{3-6}
                                         & \vspace{-0.32cm} $F_{21}(t)$	&  0.0133 &  0.0133 &  0.0182 &   0.0183 	\\ \cline{1-6}
\hline 
\end{tabular}
\end{center}
\label{table_RMSE}
\end{table}
\renewcommand*{\arraystretch}{1}
\setlength\tabcolsep{6pt} 

We also analyzed the dataset requirements of the proposed methods by considering only half of the TDS for training (we removed data randomly for the half TDS). We observed different performance for ANNs according to amounts of input data. When we trained the ANN using full TDS \emph{one-machine} technique showed better performance than the \emph{two-machines} technique, as shown in Table~\ref{table_RMSE}. This situation is reversed, however-- the \emph{two-machines} technique shows a better performance--  when we used half of the TDS for the training phase (Table~\ref{table_RMSE_half data}). Hence, the \textit{one-machine} technique performs better at modeling the channel, though it requires more data for training.
\renewcommand*{\arraystretch}{1.15}
\setlength\tabcolsep{2.5pt} 
\begin{table}[t]
\caption{Mean RMSE values for ANN techniques with half TDS}
\begin{center}
\begin{tabular}{| C{1.35cm} | C{0.8cm} | c | c || c | c |}
\cline{1-6}
\multicolumn{2}{| c |}{} & \multicolumn{2}{ c ||}{ $D=\SI{50}{\micro\meter^2/\second}$} & \multicolumn{2}{ c| }{$D=\SI{100}{\micro\meter^2/\second}$} \\ \cline{3-6}
\multicolumn{2}{| c |}{} & One ANN  &  Two ANN 	&  One ANN  &  Two ANN	\\ \hline
\vspace{-0.4cm}$d=\SI{2}{\micro\meter}$  & \vspace{-0.32cm} $F_{11}(t)$	& 0.1875 &  0.1688 &  0.3155 &    0.3222 				\\ \cline{3-6}
                                         & \vspace{-0.32cm} $F_{21}(t)$	&  0.0628 &  0.0553 &  0.0799 &    0.0729 				\\ \cline{1-6}
\vspace{-0.4cm}$d=\SI{4}{\micro\meter}$  & \vspace{-0.32cm} $F_{11}(t)$	&  0.0592 &  0.0590 &  0.0922 &    0.0926 				\\ \cline{3-6}
                                         & \vspace{-0.32cm} $F_{21}(t)$	&  0.0231 &  0.0219 &  0.0275 &    0.0270 				\\ \cline{1-6}
\vspace{-0.4cm}$d=\SI{6}{\micro\meter}$  & \vspace{-0.32cm} $F_{11}(t)$	& 0.0404 &  0.0401 &  0.0543 &    0.0540 				\\ \cline{3-6}
                                         & \vspace{-0.32cm} $F_{21}(t)$	&  0.0188 &  0.0184 &  0.0237 &   0.0229 				\\ \cline{1-6}
\vspace{-0.4cm}$d=\SI{8}{\micro\meter}$  & \vspace{-0.32cm} $F_{11}(t)$	&  0.0314 &  0.0312 &  0.0390 &    0.0388 				\\ \cline{3-6}
                                         & \vspace{-0.32cm} $F_{21}(t)$	&  0.0165 &  0.0156 &  0.0207 &    0.0205 				\\ \cline{1-6}
\vspace{-0.4cm}$d=\SI{10}{\micro\meter}$  & \vspace{-0.32cm} $F_{11}(t)$ &  0.0247 &  0.0246 &  0.0309 &   0.0306 				\\ \cline{3-6}
                                         & \vspace{-0.32cm} $F_{21}(t)$	&  0.0147 &  0.0137 &  0.0182 &   0.0188 				\\ \cline{1-6}
\hline 
\end{tabular}
\end{center}
\label{table_RMSE_half data}
\end{table}
\renewcommand*{\arraystretch}{1}
\setlength\tabcolsep{6pt} 


\subsection{Case Study: Theoretical BER Analysis}
\begin{figure}[t]
\begin{center}
	\includegraphics[width=0.9\columnwidth,keepaspectratio]%
	{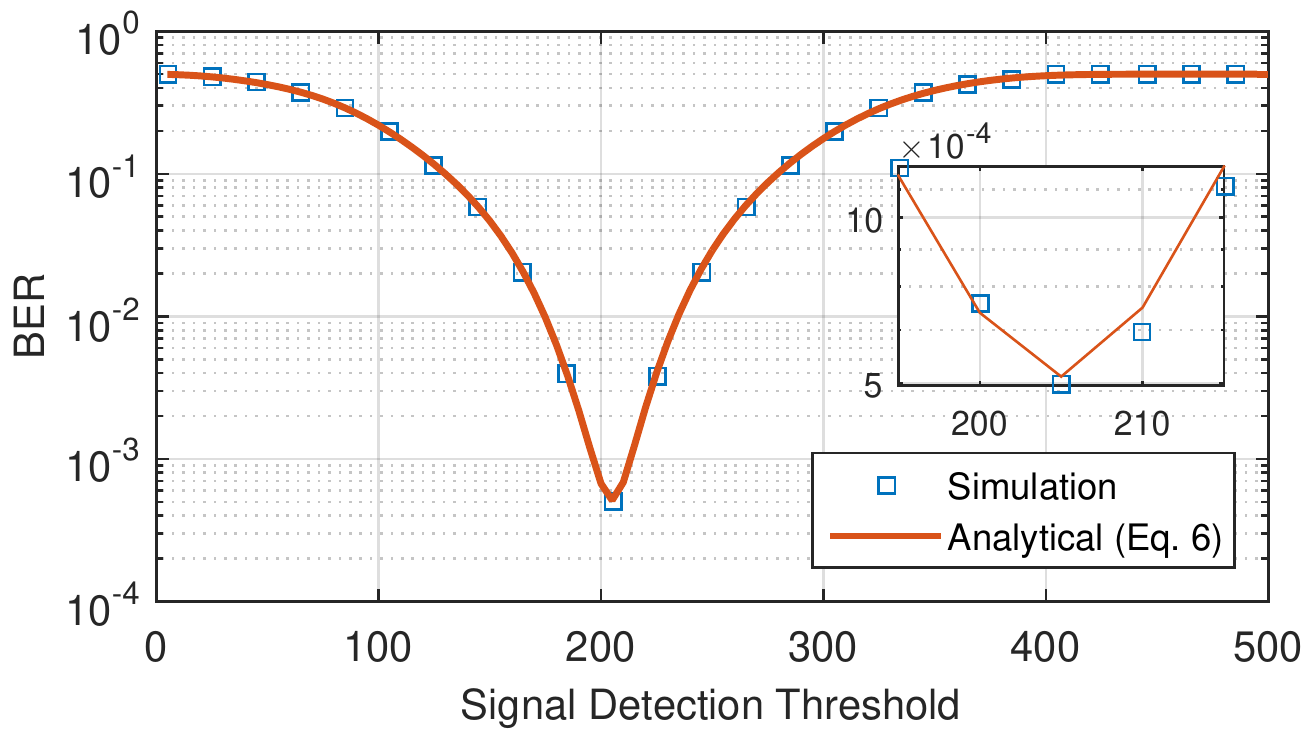}
	\caption{BER plot for $d\!=\!\SI{6}{\micro\meter}$, $h\!=\!\SI{1}{\micro\meter}$, $R\!=\!\SI{5}{\micro\meter}$, $D\!=\!\SI{100}{\micro\meter^2/\second}$, $N=1000$, $t_s=\SI{0.5}{\second}$, $\eta = 5$.}
	\label{fig____ber}
\end{center}
\end{figure}

As a case study, we considered the channel model from ANN and evaluated the BER by using \eqref{eq_ber}. To validate our model, we also ran simulations with the given parameters and obtained the empirical BER results. Simulations were replicated with consecutive $10^6$ bits. In Fig.~\ref{fig____ber}, we observe that the simulation and model results are remarkably close for the selected case from VDS. In the zoomed inset plot, we can see a small deviation from the simulation result.

\section{Conclusion}
In this work, we have developed a novel technique to model the molecular MIMO channel with two point transmitters and two absorbing spherical receivers. Currently, the literature offers no analytical model for the molecular MIMO channel in the literature yet. Therefore, we approached the problem from a novel perspective and we applied ANN-based techniques, namely \emph{one-machine} and \emph{two-machines} techniques. We showed that the proposed ANN-based techniques effectively model the molecular MIMO channel. In general, the \emph{one-machine} method performs slightly better than the \emph{two-machine} method. This performance depends, however, on the TDS size. Moreover, we showed how to utilize the channel model for theoretical BER calculations as a case study. For future work, we will focus on higher order antenna systems and TDS properties to select an appropriate ANN number.

\section*{Acknowledgment}
This work was supported by the MSIP/IITP, under the ``ICT Consilience Creative Program" (IITP-R0346-16-1008) and by the Basic Science Research Program through the NRF of Korea.


\bibliographystyle{IEEEtran}
\bibliography{mybibs_MIMO_ML}

\begin{thebibliography}{10}
\providecommand{\url}[1]{#1}
\csname url@samestyle\endcsname
\providecommand{\newblock}{\relax}
\providecommand{\bibinfo}[2]{#2}
\providecommand{\BIBentrySTDinterwordspacing}{\spaceskip=0pt\relax}
\providecommand{\BIBentryALTinterwordstretchfactor}{4}
\providecommand{\BIBentryALTinterwordspacing}{\spaceskip=\fontdimen2\font plus
\BIBentryALTinterwordstretchfactor\fontdimen3\font minus
  \fontdimen4\font\relax}
\providecommand{\BIBforeignlanguage}[2]{{%
\expandafter\ifx\csname l@#1\endcsname\relax
\typeout{** WARNING: IEEEtran.bst: No hyphenation pattern has been}%
\typeout{** loaded for the language `#1'. Using the pattern for}%
\typeout{** the default language instead.}%
\else
\language=\csname l@#1\endcsname
\fi
#2}}
\providecommand{\BIBdecl}{\relax}
\BIBdecl

\bibitem{farsad2016comprehensiveSO}
N.~Farsad, H.~B. Yilmaz, A.~W. Eckford, C.-B. Chae, and W.~Guo, ``A
  comprehensive survey of recent advancements in molecular communication,''
  \emph{{IEEE} Commun. Surveys Tuts.}, vol.~18, no.~3, pp. 1887--1919, 2016.

\bibitem{nakano2013molecularC_BOOK}
T.~Nakano, A.~W. Eckford, and T.~Haraguchi, \emph{Molecular
  communication}.\hskip 1em plus 0.5em minus 0.4em\relax Cambridge University
  Press, 2013.

\bibitem{akyildiz2011nanonetworksAN}
I.~F. Akyildiz, J.~M. Jornet, and M.~Pierobon, ``Nanonetworks: a new frontier
  in communications,'' \emph{Commun. {ACM}}, vol.~54, no.~11, pp. 84--89, Nov.
  2011.

\bibitem{guo2015molecularVE}
W.~Guo, C.~Mias, N.~Farsad, and J.-L. Wu, ``Molecular versus electromagnetic
  wave propagation loss in macro-scale environments,'' \emph{{IEEE} Trans. Mol.
  Bio. Multi-Scale Commun.}, vol.~1, no.~1, pp. 18--25, 2015.

\bibitem{arjmandi2013diffusionBN}
H.~Arjmandi, A.~Gohari, M.~N. Kenari, and F.~Bateni, ``Diffusion-based
  nanonetworking: A new modulation technique and performance analysis,''
  \emph{{IEEE} Commun. Lett.}, vol.~17, no.~4, pp. 645--648, Apr. 2013.

\bibitem{noe2014optimalRD}
A.~Noel, K.~Cheung, and R.~Schober, ``Optimal receiver design for diffusive
  molecular communication with flow and additive noise,'' \emph{{IEEE} Trans.
  NanoBiosci.}, vol.~13, no.~3, pp. 350--362, Sep. 2014.

\bibitem{hsieh2013asynchronousIE}
Y.-P. Hsieh, Y.-C. Lee, P.-J. Shih, P.-C. Yeh, and K.-C. Chen, ``On the
  asynchronous information embedding for event-driven systems in molecular
  communications,'' \emph{Elsevier Nano Commun. Netw.}, vol.~4, no.~1, pp. 2 --
  13, Mar. 2013.

\bibitem{kilinc2013receiverDF_JSAC}
D.~Kilinc and O.~B. Akan, ``Receiver design for molecular communication,''
  \emph{{IEEE} J. Sel. Areas Commun.}, vol.~31, no.~12, pp. 705--714, 2013.

\bibitem{azadi2016novelEM}
M.~Azadi and J.~Abouei, ``A novel electrical model for
  advection-diffusion-based molecular communication in nanonetworks,''
  \emph{{IEEE} Trans. NanoBiosci.}, vol.~15, no.~3, pp. 246--257, 2016.

\bibitem{srinivas2012molecularCI_inverseG}
K.~V. Srinivas, A.~W. Eckford, and R.~S. Adve, ``Molecular communication in
  fluid media: The additive inverse {Gaussian} noise channel,'' \emph{{IEEE}
  Trans. Inf. Theory}, vol.~58, no.~7, pp. 4678--4692, Jul. 2012.

\bibitem{kadlor2012molecularCU_drift_TNBS}
S.~Kadloor, R.~S. Adve, and A.~W. Eckford, ``Molecular communication using
  {Brownian} motion with drift,'' \emph{{IEEE} Trans. NanoBiosci.}, vol.~11,
  no.~2, pp. 89--99, June 2012.

\bibitem{yilmaz2014threeDC}
H.~B. Yilmaz, A.~C. Heren, T.~Tugcu, and C.-B. Chae, ``Three-dimensional
  channel characteristics for molecular communications with an absorbing
  receiver,'' \emph{{IEEE} Commun. Lett.}, vol.~18, no.~6, pp. 929--932, Jun.
  2014.

\bibitem{akkaya2015effectOR_receptor_COML}
A.~Akkaya, H.~B. Yilmaz, C.-B. Chae, and T.~Tugcu, ``Effect of receptor density
  and size on signal reception in molecular communication via diffusion with an
  absorbing receiver,'' \emph{{IEEE} Commun. Lett.}, vol.~19, no.~2, pp.
  155--158, Feb. 2015.

\bibitem{koo2016molecularMIMO_JSAC}
B.-H. Koo, C.~Lee, H.~B. Yilmaz, N.~Farsad, A.~Eckford, and C.-B. Chae,
  ``Molecular {MIMO}: From theory to prototype,'' \emph{{IEEE} J. Sel. Areas
  Commun.}, vol.~34, no.~3, pp. 600--614, Mar. 2016.

\bibitem{kuran2011modulationTF_ICC}
M.~S. Kuran, H.~B. Yilmaz, T.~Tugcu, and I.~F. Akyildiz, ``Modulation
  techniques for communication via diffusion in nanonetworks,'' in \emph{Proc.
  IEEE Int. Conf. on Commun. (ICC)}, Jun. 2011, pp. 1--5.

\bibitem{nr13}
N.-R. Kim and C.-B. Chae, ``Novel modulation techniques using isomers as
  messenger molecules for nano communication networks via diffusion,''
  \emph{{IEEE} J. Sel. Areas Commun.}, vol.~31, no.~12, pp. 847--856, Dec.
  2012.

\end{thebibliography}

\end{document}